\RequirePackage{fix-cm}
\documentclass[smallextended]{svjour3}       
\smartqed  
\usepackage{graphicx}

\usepackage{amsmath,amsthm,amsfonts}
\usepackage{amssymb,mathrsfs}
\usepackage{hyperref}
\usepackage{float}
\usepackage{relsize}
\usepackage{siunitx}
\usepackage{geometry}
\usepackage{dcolumn}
\usepackage{bm}
\newcommand{\lambdabar}{{\mkern0.75mu\mathchar '26\mkern -9.75mu\lambda}}
\begin{document}

\title{The gyroscopic frequency of metric $f(R)$ and generalised Brans--Dicke theories: constraints from Gravity Probe--B}


\author{A. Dass        \and
        S. Liberati 
}


\institute{A. Dass \at
               Dipartimento di Fisica, Universit\`a degli Studi di Trento, Via Sommarive, 14, 38123 Povo, Trento TN, Italy and SISSA, Italy \\
              \email{abhinandan.dass@alumni.unitn.it}           
           \and
           S. Liberati \at
              SISSA, Via Bonomea, 265, 34136 Trieste TS, Italy and INFN, sezione di Trieste, Italy.\\
              \email{liberati@sissa.it}
}

\date{Received: date / Accepted: date}

\maketitle

\begin{abstract}
We confront the predicted gyroscopic precession (in particular the geodetic precession) from metric $f(R)$ theory with the data provided by the mission, Gravity Probe--B. We find the constraint, $|a_2| < 1.33\times 10^{12} \mathrm{m}^2$, where $a_2$ is the coefficient assessing the strength of the lowest order correction to the Einstein--Hilbert action for a metric $f(R)$ theory with $f$ analytic. This constraint improves over astrophysical bounds provided by massive black holes and planetary precession which are $|a_2|\gtrsim 10^{17} \si{m^2}$ and $|a_2|\lesssim 1.2\times10^{18} \si{m^2}$ respectively and it is complementary to the stringent ones provided by lab based experiments, like the E\"ot--Wash experiment. We also investigate the modification of our result for gyroscopic precession if the oblateness of Earth is taken into account by considering the quadrupole moment of Earth. Finally, we provide a generalisation of  our result for the gyroscopic precession in the context of Brans--Dicke theories with a potential (recovering the previously derived results in the appropriate limits).
\end{abstract}

\section{Linearised Metric $f(R)$ gravity}\label{1}
Accurate experiments probing the Earth gravitational field are providing a new venue to test deviations from General Relativity (GR) predictions. In particular $f(R)$ has been used to model this deviations for it being the most natural extension of the Einstein--Hilbert Lagrangian. For an instance, a particular $f(R)$ Lagrangian was derived as the effective classical Lagrangian leading to the modified Friedmann equations of Loop Quantum Cosmology, both with a metric and a Palatini ansatz \cite{sotiriou2009covariant,olmo2009covariant}. Berry et al looked at linearised $f(R)$ to impose constraints from planetary precession and gravitational--wave astronomy \cite{berry2011linearized}. In this particular work, we consider a metric $f(R)$ with $f$ analytic and find constraints from the Gravity Probe--B (GP--B) measurements of the geodetic precession.\\\\

We can extend General Relativity (GR) by higher than second order field equations. This can improve renormalisability properties by  allowing the graviton propagator to fall off more quickly in the UV regime. However, it can introduce ghost degrees of freedom causing instabilities \cite{woodard2007avoiding}.\\\\ In $f(R)$ theories, a generic function of the Ricci scalar is employed instead of the usual linear term. Here, a general function of the Ricci scalar which leads to fourth--order field equations is considered. Such theories have improved renormalisation properties \cite{stelle1977renormalization} without ghost and could also possibly provide an inflationary phase \cite{starobinsky1980new}.\\\\
In the present analysis, a linearised metric $f(R)$ theory where $f(R)$ is taken to be analytic about $R=0$ is considered. The $(-+++)$ space--like convention is used and the d'alembertian is defined as $\square=-g^{\mu\nu}\nabla_{\mu}\nabla_{\nu}$ and prime denotes differentiation with respect to $R$. \\\\
The choice of $f(R)$ to be Taylor expandable about $R=0$ is made because of the following reasons \cite{dass_liberati2}:
\begin{itemize}
\item It is found that $\dfrac{1}{R}$ models do not seem to have the correct Newtonian limit and there is no considerable evidence that they pass the solar system tests \cite{sotiriou2007metric}.
\item It can be shown that $\dfrac{1}{R}$ models lead to instability in the weak gravity regime \cite{dolgov2003can}. 
\end{itemize}
A metric theory is chosen over the Palatini one because in this case, even a simple polytropic equation of state leads to a curvature singularity for a static spherically symmetric solution \cite{dolgov2003can}. Moreover, as we will see later, the Palatini theory can be shown to be classically equivalent to a singular generalized Brans--Dicke theory. \\\\
The linearisation procedure given in \cite{berry2011linearized} is followed. As such, an analytic $f(R)$ can be expanded around $R=0$ as 
\begin{equation}
f(R)=a_0+a_1R+\frac{a_2}{2!}R^2+\frac{a_3}{3!}R^3+\hdots
\end{equation}
As the dimension of $f(R)$ has to be the same as that of $R$. We have, $[a_{n}]=[R]^{(1-n)}$. Also the requirement of correct GR limit tells us that $a_{1}=1$, any rescaling will be included in the definition of $G$.\\\\The vacuum field equations can be written as
\begin{equation}
f^\prime R_{\mu \nu} - \nabla_\mu \nabla_\nu f^\prime + g_{\mu \nu}\square f^\prime - \frac{f}{2}g_{\mu \nu} = 0.
\end{equation}
Tracing the above equation gives us
\begin{equation}\label{1.105}
f^\prime R + 3\square f^\prime -2f = 0.
\end{equation}
Note that for a uniform flat spacetime, $R=0$, which gives \cite{capozziello2007newtonian}
\begin{equation}\label{1.106}
a_0=0,
\end{equation}
which tantamounts to saying that for such solution to exist the cosmological constant cannot be present.\\\\ In analogy with the Einstein tensor of GR, we define the following
\begin{equation}\label{1.107}
\mathcal{G}_{\mu \nu} \equiv f^\prime R_{\mu \nu} - \nabla_\mu \nabla_\nu f^\prime + g_{\mu \nu}\square f^\prime - \frac{f}{2}g_{\mu \nu},
\end{equation}
such that in vacuum, we have
\begin{equation}
\mathcal{G}_{\mu \nu}=0.
\end{equation}
We are interested in considering the case of a perturbed metric about a Minkowski background, $g_{\mu\nu}=\eta_{\mu\nu}+h_{\mu\nu}$, similar to standard analysis done in GR. The linearised connection is found to be
\begin{equation}
\Gamma^{(1)^\rho}_{\mu \nu}=\frac{1}{2}\eta^{\rho \lambda}(\partial_\mu h_{\lambda \nu} + \partial_\nu h_{\lambda \mu} - \partial_\lambda h_{\mu \nu}),
\end{equation}
while the linearised Riemann tensor is given by
\begin{equation}
R^{(1)^\lambda}_{\mu \nu \rho}=\frac{1}{2}(\partial_\mu \partial_\nu h^\lambda_\rho + \partial^\lambda \partial_\rho h_{\mu \nu} - \partial_\mu \partial_\rho h^\lambda_\nu - \partial^\lambda \partial_\nu h_{\mu \rho}),
\end{equation}
where as usual the flat metric is used to raise and lower the indices.\\\\
The linearised Ricci tensor is obtained by the self contraction of the above equation to give
\begin{equation}\label{1.111}
R_{\mu\nu}^{(1)}=\frac{1}{2}(\partial_\mu \partial_\rho h^\rho_\nu+\partial_\nu \partial_\rho h^\rho_\mu-\partial_\mu \partial_\nu h-\Box h_{\mu\nu}),
\end{equation}
while a further contraction with the flat metric gives the Ricci scalar
\begin{equation}\label{1.112}
R^{(1)}=\partial_\mu \partial_\rho h^{\rho \mu} - \square h.
\end{equation}
To the first order in $h_{\mu\nu}$, we can express $f(R)$ as a Maclaurin series
\begin{equation}\label{377}
f(R)=a_0+R^{(1)},
\end{equation}
\begin{equation}\label{378}
f^\prime=1+a_2R^{(1)},
\end{equation}
Perturbing around a Minkowski background where the Ricci scalar vanishes, we make use of equation \eqref{1.106} to set $a_0=0$ and insert the resulting equations in \eqref{1.107}
\begin{equation}\label{1.115}
\mathcal{G}^{(1)}_{\mu \nu} = R^{(1)}_{\mu\nu}-\partial_\mu \partial_\nu(a_2R^{(1)}) + \eta_{\mu \nu}\square(a_2R^{(1)}) - \frac{R^{(1)}}{2}\eta_{\mu \nu}.
\end{equation}
While from the linearised trace equation, \eqref{1.105}, we get
\begin{equation}\label{1.116}
\mathcal{G}^{(1)} = 3\square(a_2R^{(1)})-R^{(1)},
\end{equation}
where $\mathcal{G}^{(1)}=\eta^{\mu\nu}\mathcal{G}_{\mu\nu}^{(1)}$. We observe that this is the massive inhomogeneous Klein--Gordon equation. Setting $\mathcal{G}=0$ for a vacuum solution in $f(R)$ at all orders of perturbations, the standard Klein--Gordon equation is obtained
\begin{equation}\label{576}
\square R^{(1)} + \Upsilon^2 R^{(1)} = 0.
\end{equation}
Where a reciprocal length is defined as 
\begin{equation}\label{75}
\Upsilon^2 = -\frac{1}{3a_2}.
\end{equation}
For physically meaningful solution, we require that $\Upsilon^2>0$ and hence we constrain $f(R)$ such that $a_2<0$ \cite{schmidt1986h,teyssandier1990new,olmo2005gravity,corda2008massive}. From $\Upsilon$, a reduced Compton wavelength associated with the scalar mode is defined \cite{berry2011linearized}. 
\begin{equation}
\lambdabar=\frac{1}{\Upsilon}.
\end{equation}
In order to look for wave solutions, the linearized Einstein tensor and its trace is expresssed in terms of the perturbation of the background flat geometry, $h_{\mu\nu}$ and $h$. As a consequence, a quantity $\bar{h}_{\mu\nu}$ is needed that will satisfy a wave equation and is related to $h_{\mu\nu}$ as
\begin{equation}
\bar{h}_{\mu \nu} = h_{\mu \nu} + A_{\mu \nu}.
\end{equation}
One normally uses the trace-reversed form in GR, where $A_{\mu\nu}=-\dfrac{h}{2}\eta_{\mu\nu}$. However, it is evident that this will not be sufficient in this case for a wave solution. Hence, we shall look for a solution on similar lines by introducing the ansatz
\begin{equation}\label{1.120}
\bar{h}_{\mu \nu} = h_{\mu \nu} - \frac{h}{2}\eta_{\mu \nu} + B_{\mu \nu},
\end{equation}
where $B_{\mu\nu}$ is a symmetric rank--2 tensor. The only rank-two tensors in our theory are $h_{\mu\nu}$, $\eta_{\mu\nu}$, $R_{\mu\nu}^{(1)}$, and $\partial_{\mu}\partial_{\nu}$. We notice that $B_{\mu\nu}$ needs to be first order in
$h$, and depends on $f(R)$. This can be easily accomplished by the following ansatz \cite{berry2011linearized}
\begin{equation}\label{1.121}
\bar{h}_{\mu \nu} = h_{\mu \nu} + \Big(a_2bR^{(1)}-\frac{h}{2}\Big)\eta_{\mu \nu},
\end{equation}
where $a_2$ has been introduced for dimensional consistency and $b$ is a dimensionless number. The contraction with the flat metric gives
\begin{equation}
\bar{h}=4a_2bR^{(1)}-h.
\end{equation}
We eliminate $h$ in \eqref{1.121} to obtain
\begin{equation}\label{76}
h_{\mu \nu} = \bar{h}_{\mu \nu} + \Big(a_2bR^{(1)}-\frac{\bar{h}}{2}\Big)\eta_{\mu \nu}.
\end{equation}
Similar to GR, we have the freedom to perform a gauge transformation \cite{weinberg2014gravitation} given that the field equations are gauge invariant (Since the Lagrangian is a function of gauge invariant Ricci scalar). Following the usual treatment in GR, a de Donder gauge is assumed
\begin{equation}
\nabla^\mu \bar{h}_{\mu \nu} = 0,
\end{equation}
which in flat spacetime gives
\begin{equation}
\partial^\mu \bar{h}_{\mu \nu} = 0.
\end{equation}
Subject to the above conditions, the Ricci tensor \eqref{1.111} becomes
\begin{equation}
R^{(1)}_{\mu \nu} = -\frac{1}{2} \Big[ 2b\partial_\mu \partial_\nu (a_2R^{(1)}) + \square \Big(\bar{h}_{\mu \nu} - \frac{\bar{h}}{2}\eta_{\mu \nu} \Big)+ \frac{b}{3}(R^{(1)} + \mathcal{G}^{(1)})\eta_{\mu \nu} \Big].
\end{equation}
Contraction of the above results in
\begin{equation}
R^{(1)} = -\frac{1}{2}\Big[2b\Box(a_2R^{(1)})-\Box \bar{h}+\frac{4b}{3}(R^{(1)}+G^{(1)})\Big].
\end{equation}
We replace $G^{(1)}$ above by equation \eqref{1.116}, the Ricci scalar becomes
\begin{equation}\label{753}
R^{(1)} = -3b\Box(a_2R^{(1)})+\frac{1}{2}\Box\bar{h}.
\end{equation}
The above expression in \eqref{1.115} is used to give
\begin{equation}\label{1.128}
\mathcal{G}^{(1)}_{\mu \nu} = \frac{2-b}{6}\mathcal{G}^{(1)}\eta_{\mu \nu} - \frac{1}{2}\square\Big(\bar{h}_{\mu \nu} - \frac{\bar{h}}{2}\eta_{\mu \nu} \Big)- (b+1)\Big[\partial_\mu \partial_\nu (a_2R^{(1)}) + \frac{1}{6}R^{(1)}\eta_{\mu \nu} \Big].
\end{equation}
Without a loss of generality, fixing $b=-1$, so that the last term above vanishes and equations \eqref{1.121} and \eqref{76} becomes \cite{corda2008massive,capozziello2008massive}
\begin{equation}
\bar{h}_{\mu \nu} = h_{\mu \nu} - \Big(a_2R^{(1)}+\frac{h}{2}\Big)\eta_{\mu \nu},
\end{equation}
\begin{equation}
h_{\mu \nu} = \bar{h}_{\mu \nu} - \Big(a_2R^{(1)}+\frac{\bar{h}}{2}\Big)\eta_{\mu \nu}.
\end{equation}
From \eqref{753}, the Ricci scalar becomes
\begin{equation}
R^{(1)} = 3\square(a_2R^{(1)}) + \frac{1}{2}\square\bar{h}.
\end{equation}
To be consistent with \eqref{1.116}, we require that
\begin{equation}\label{659}
-\frac{1}{2}\square\bar{h}=\mathcal{G}^{(1)}.
\end{equation}
Inserting the above expression in \eqref{1.128} along with $b=-1$, we obtain
\begin{equation}\label{660}
-\frac{1}{2}\square\bar{h}_{\mu \nu}=\mathcal{G}^{(1)}_{\mu \nu}.
\end{equation}
If $a_2$ is sufficiently small such that it can be neglected, the equations \eqref{659} and \eqref{660} drops down to that of GR.\\\\
Adding a source term $T_{\mu\nu}$, the linearised equations are found at the first order in perturbation theory
\begin{equation}\label{499}
-{\frac{1}{2}}\Box \bar {h}=G^{(1)}={\frac{8 \pi G}{c^4}}T,
\end{equation}
\begin{equation}\label{456}
-{\frac{1}{2}}\Box \bar {h}_{\mu \nu}=G^{(1)}_{\mu\nu}={\frac{8 \pi G}{c^4}}T_{\mu \nu},
\end{equation}
which we rewrite as
\begin{equation}\label{35}
\Box \bar {h}_{\mu \nu}=-{\frac{16 \pi G}{C^4}}T_{\mu \nu},
\end{equation}
which is the tensor mode wave equation.
For the scalar mode, using \eqref{499} in \eqref{1.116} and remembering \eqref{75}, the following equation needs to be solved
\begin{equation}\label{34}
\Box  R^{(1)} + \Upsilon ^2 R^{(1)}={\frac{8 \pi G}{c^4}}\Upsilon^2 T.
\end{equation}
To solve the above two equations, \eqref{35} and \eqref{34} with the source, the following Green function is introduced
\begin{equation}
(\Box +\Upsilon^2)\mathscr{G}_\Upsilon (x,x')=\delta(x-x'),
\end{equation}
where $\Box$ acts on $x$, and $\mathscr{G}_\Upsilon$ is given by
\begin{equation}
\mathscr{G}_\Upsilon (x,x')=\frac{1}{(2\pi)^4}\int d^4p\frac{\exp[-ip\cdot(x-x')]}{\Upsilon^2-p^2}.
\end{equation}
A contour integration method is used to solve the above to give \cite{peskin2018introduction}
\begin{equation}\label{2.81}
\mathscr{G}_\Upsilon (x,x')=
\begin{cases}
\mathlarger{\int}{\dfrac{d\omega}{2\pi}}\exp[-i\omega(t-t')]{\frac{1}{4\pi r}}\exp[i(\omega^2 - \Upsilon^2)^{{\frac{1}{2}}}r], \quad \omega^2>\Upsilon^2\\
\mathlarger{\int}{\dfrac{d\omega}{2\pi}}\exp[-i\omega(t-t')]{\frac{1}{4\pi r}}\exp[-(\Upsilon^2-\omega^2 )^{{\frac{1}{2}}}r],\quad \omega^2<\Upsilon^2\\
\end{cases}
\end{equation}
where we have, $t=x^{0}$, $t'=x'^{0}$ and $r=|x-x'|$.\\\\ We see that the tensor equation \eqref{35} does not have an
associated mass (the graviton is still massless).
Indeed, asking for the above expressions to be a solution of \eqref{35} shows that the relevant Green function are those with $\Upsilon=0$
\begin{equation}
\mathscr{G}_{0} (x,x')=\frac{\delta(t-t'-r)}{4\pi r},
\end{equation}
which is the retarded time Green function. Using it to solve \eqref{35}, one obtains
\begin{equation*}
\bar{h}_{\mu\nu}=-{\frac{16\pi G}{c^4}}\int d^4x' \mathscr{G}_{0} (x,x') T_{\mu\nu}(x')
\end{equation*}
\begin{equation}\label{2.79}
 \bar {h}_{\mu \nu}=-{\frac{4G}{c^4}}\int d^3x' {\frac{T_{\mu \nu}(t-r,x')}{r}}.
\end{equation}
The scalar mode equation \eqref{34} can be solved as
\begin{equation}\label{692}
R^{(1)}(x)=-8\pi G\Upsilon ^2\int d^4x' \mathscr{G}_\Upsilon (x,x') T(x'),
\end{equation}
Going to the Newtonian limit and considering a stationary, point source mass distribution. We have
\begin{align}\label{2.82}
T_{00} &=\rho c^2, & T_{0i} &=-cj_i, & \rho &=M\delta (x'), &T_{ij} \approx 0,
\end{align}
where $\vec{j}=\rho\vec{v}$ is the mass current.
And 
\begin{align*}
|T_{00}| & \gg |T_{0i}|, & |T_{00}|\gg|T_{ij}|.
\end{align*}
So we get,
\begin{equation}\label{2.83}
 \bar {h}_{00}=-{\frac{4GM}{r c^2}}.
\end{equation}
 In order to find $R^{(1)}$, we define in the equation \eqref{2.81}
\begin{equation}\label{678}
f(r,\omega)=
\begin{cases}
\exp[i(\omega^2 - \Upsilon^2)^{{\frac{1}{2}}}r],\quad \omega^2>\Upsilon^2\\
\exp[-(\Upsilon^2-\omega^2 )^{{\frac{1}{2}}}r],\quad \omega^2<\Upsilon^2\\
\end{cases}
\end{equation}
Then from equation \eqref{692}, we obtain
\begin{align}\label{2.85}
R^{(1)}(x) &=-8\pi G\Upsilon ^2\int d^4x' \mathscr{G}_\Upsilon (x,x') M\delta^{3}(x')\nonumber \\
&=-8\pi G\Upsilon ^2\int dt'\int {\frac{d\omega}{2\pi}}\exp[-i\omega(t-t')]M{\frac{1}{4\pi r}}f(r,\omega)\\
&=-8\pi G\Upsilon ^2 M{\frac{1}{4\pi r}}f(r,0)\nonumber \\
&=-2 G\Upsilon^2 M{\frac{\exp(-\Upsilon r)}{r}}\nonumber .
\end{align}
Using $\bar {h}=\bar {h}_{00}$, \eqref{75}, \eqref{2.83} and  \eqref{2.85} in \eqref{76}, we obtain
\begin{equation}\label{339}
h_{00}=-{\frac{2GM}{r}}\left[1 + \frac {\exp(-\Upsilon r)}{3}\right].
\end{equation}
In a similar manner, we can show that the purely space component has the following form
\begin{equation}\label{1056}
h_{ij}=-{\frac{2GM}{r}}\left[1 - \frac {\exp(-\Upsilon r)}{3}\right]\delta_{ij}.
\end{equation}
In addition to this, we can extend the result and consider a slowly rotating source with angular momentun $\vec{J}$, then we have an additional term $\bar {h}^{0i}=h^{0i}$ \cite{berry2011linearized,hobson2006general}.\\\\
\begin{equation}\label{2981}
\bar {h}^{0i}=h^{0i}=-\frac{2G}{c^3r^3} (\vec J \times \vec r)_i.
\end{equation}
Now, we can define as usual $\bar {h}_{00}$ as
\begin{equation}\label{2.93}
\bar {h}_{00}\equiv -\frac{4\Phi}{c^2},
\end{equation}
where $\Phi$ is the scalar potential. Various authors have used different definitions for this potential. While a few have taken it with an overall $+$ sign like in \cite{hobson2006general}, some uses the definition with a $-$ sign as in \eqref{2.93} \cite{padmanabhan2010gravitation}.\\\\
In a similar manner, we can define $\bar {h}^{0i}$ as \cite{hobson2006general,padmanabhan2010gravitation}
\begin{equation}\label{5891}
\bar {h}^{0i}=h^{0i}\equiv\frac{2A_i}{c^2},
\end{equation}
where $A_i$ is the vector potential.\\\\
Taking into account the expression of the metric perturbation for purely time component \eqref{339}, purely space component \eqref{1056}, and \eqref{5891}, we get for the line element in metric $f(R)$
\begin{equation}
ds^2=  -c^2\left\{{1-\frac{2\Phi}{c^2} \left[1+ \frac {\exp(-\Upsilon r)}{3}\right]}\right\} dt^2 - \frac{4}{c}(\vec{A} \cdot d\vec{r})dt + \left\{{1+\frac{2\Phi}{c^2} \left[1- \frac {\exp(-\Upsilon r)}{3}\right]}\right\} dr^2,
\end{equation}
where $\Phi=\dfrac{GM}{r}$ is the scalar potential.\\\\
Let us now define the following symbols
\begin{equation}\label{2.110}
\alpha \equiv \Phi\left[1+ \frac {\exp(-\Upsilon r)}{3}\right],
\end{equation}
\begin{equation}\label{2.111}
\beta \equiv \Phi\left[1- \frac {\exp(-\Upsilon r)}{3}\right],
\end{equation}
so as to write the line element in the following concise form
\begin{equation}\label{2.115}
ds^2 =-c^2\left(1-\frac{2\alpha}{c^2}\right)dt^2 -\frac{4}{c}\left(\vec{A} \cdot d\vec{r}\right)dt + \left(1+\frac{2\beta}{c^2}\right)dr^2,
\end{equation}
\section{Post--Newtonian Approximation}
To derive the expression for gyroscopic precession frequency, we need to resort to post-Newtonian approximation given that it is a higher order effect \cite{weinberg2014gravitation}. So, in considering a motion of a particle in the above metric, we shall consider terms up to order $\dfrac{\vec{v}^4}{r}$ where $\vec{v}$ is the velocity and $r$ is the radial distance of the particle. We will derive the complete precession frequency and then later, we will make the distinction between the Lense--Thirring part and the geodetic part.\\\\
The equation of motion of a particle is given by the so called geodesic equation. 
\begin{equation}
\frac{d^2x^{\mu}}{d\tau^2}+\Gamma_{\nu\lambda}^{\mu} \frac{dx^{\nu}}{d\tau}\frac{dx^{\lambda}}{d\tau}=0,
\end{equation}
where $\Gamma_{\nu\lambda}^{\mu}$ are the affine or the Levi--Civita coefficients and $\tau$ is the proper time.
From the geodesic equation and by employing the product rule of derivatives, we can compute the acceleration of the particle trivially as \cite{weinberg2014gravitation}
\begin{align}
\frac{d^2x^{i}}{dt^2} &=\left(\frac{dt}{d\tau}\right)^{-1}\frac{d}{d\tau}\left[\left(\frac{dt}{d\tau}\right)^{-1}\frac{dx^i}{d\tau}\right]\\
&=\left(\frac{dt}{d\tau}\right)^{-2}\frac{d^2x^i}{d\tau^2}-\left(\frac{dt}{d\tau}\right)^{-3}\frac{d^2t}{d\tau^2}\frac{dx^i}{d\tau}\\
&= -\Gamma_{\nu\lambda}^i \frac{dx^{\nu}}{dt}\frac{dx^{\lambda}}{dt}+\Gamma_{\nu\lambda}^{0}\frac{dx^{\nu}}{dt}\frac{dx^{\lambda}}{dt}\frac{dx^i}{dt},
\end{align}
where $dt$ is the time. This can be further expounded as 
\begin{equation}\label{3.5}
\frac{d^2x^i}{dt^2}= -\Gamma_{00}^{i}-2\Gamma_{0j}^{i}\frac{dx^j}{dt}-\Gamma_{jk}^{i}\frac{dx^j}{dt}\frac{dx^k}{dt}  +\left[\Gamma_{00}^{0} + 2\Gamma_{0j}^{0}\frac{dx^j}{dt}+\Gamma_{jk}^{0}\frac{dx^j}{dt}\frac{dx^k}{dt}\right]\frac{dx^i}{dt}.
\end{equation}
In Newtonian approximation, we treat the velocities to be vanishingly small and we keep only terms of first order in the difference between the $g_{\mu\nu}$ and $\eta_{\mu\nu}$. So, we get for the acceleration
\begin{equation}
\frac{d^2x^i}{dt^2}\approx -\Gamma_{00}^{i}=\frac{1}{2}\frac{\partial g_{00}}{\partial x^i}.
\end{equation}
But $g_{00}-1$ is of order $\dfrac{GM}{r}$. So, $\dfrac{d^2x^i}{dt^2}\sim\dfrac{v^2}{r}$. In post-Newtonian approximation, we want to determine instead that $\dfrac{d^2x^i}{dt^2}$ up to order $\dfrac{\vec{v}^4}{r}$.\\\\
Hence, for finding the acceleration, \eqref{3.5}, in the regime of post-Newtonian approximation ($\sim\dfrac{\vec{v}^4}{r}$), we need the following components of the connection\\\\
$\Gamma_{00}^{i}$ up to order $\dfrac{\vec{v}^4}{r}$\\
$\Gamma_{0j}^{i}$ and $\Gamma_{00}^0$ upto order $\dfrac{\vec{v}^3}{r}$\\
$\Gamma_{jk}^{i}$ and $\Gamma_{0j}^{0}$  upto order $\dfrac{\vec{v}^2}{r}$\\
$\Gamma_{jk}^{0}$ up to order $\dfrac{\vec{v}}{r}$\\\\
So, let us introduce the symbol $\overset{N}{\Gamma_{\mu \nu}^\lambda}$ to represent $\Gamma_{\mu \nu}^\lambda$ up to order $\dfrac{\vec{v}^N}{r}$.\\\\
We have
\begin{equation}\label{3.7}
\overset{3}{\Gamma_{i0}^j}=\frac{1}{2}\left[\overset{3}{\frac{\partial g_{i0}}{\partial x^j}}+\overset{2}{\frac{\partial g_{ij}}{\partial t}}-\overset{3}{\frac{\partial g_{j0}}{\partial x^i}}\right],
\end{equation}
\begin{equation}\label{3.8}
\overset{2}{\Gamma_{i0}^0}=-\frac{1}{2}\overset{2}{\frac{\partial g_{00}}{\partial x^i}},
\end{equation}
\begin{equation}\label{3.9}
\overset{2}{\Gamma_{ik}^j}=\frac{1}{2}\left[\overset{2}{\frac{\partial g_{ij}}{\partial x^k}}+\overset{2}{\frac{\partial g_{ik}}{\partial x^j}}-\overset{2}{\frac{\partial g_{jk}}{\partial x^i}}\right],
\end{equation}
where the superscript number over the metric derivatives also denotes the $\dfrac{\vec{v}^N}{r}$ behaviour.
\subsection{\textbf{Gyroscope Precession}}
A free falling particle will have a four-velocity, $U^{\nu}=\dfrac{dx^{\nu}}{d\tau}$ and Spin $S_{\mu}$. From principle of general covariance, the spin of a particle in free fall precesses according to the parallel transport equation \cite{weinberg2014gravitation} which is given by
\begin{equation}\label{3.10}
\frac{dS_{\mu}}{d\tau}=\Gamma_{\mu\nu}^{\lambda} S_{\lambda} \frac{dx^{\nu}}{d\tau}.
\end{equation}
Also, $S_{\mu}$ is orthogonal to the velocity and hence, we have
\begin{equation}\label{3.11}
\frac{dx^{\mu}}{d\tau}S_{\mu}=0.
\end{equation}
Or in other words, if we resolve it into time and space components, we can write the following
\begin{equation}\label{3.12}
S_0=-v^i S_i.
\end{equation}
We set $\mu=i$ in equation \eqref{3.10} and multiply by $\dfrac{d\tau}{dt}$ and use \eqref{3.12} to eliminate $S_0$, which then gives us the following expression 
\begin{equation}\label{33}
\frac{dS_{i}}{dt}=\Gamma_{i0}^{j}S_j-\Gamma_{i0}^{0}v^jS_j+\Gamma_{ik}^{j}v^kS_j-\Gamma_{ik}^{0}v^kv^jS_j.
\end{equation}
Now, if we observe the above expression, we will realise that post--Newtonian approximation allows us to evaluate coefficients of $S_j$ on the right hand side of the above equation to order $\dfrac{\vec{v}^3}{r}$, which then gives us
\begin{equation}\label{3.14}
\frac{dS_{i}}{dt}\approx[\overset{3}{\Gamma_{i0}^j}-\overset{2}{\Gamma_{i0}^0}v^j+\overset{2}{\Gamma_{ik}^j}v^k]S_j.
\end{equation}
The last term in \eqref{33} drops out because the term, $\overset{1}{\Gamma_{ik}^0}$, is not present in the post-Newtonian approximation as we have already discussed in the preceding section, i.e., equations \eqref{3.7}, \eqref{3.8}, \eqref{3.9}. \\\\
To calculate \eqref{3.14}, let us recall the particular metric solution of $f(R)$ that we discussed previously, i.e. Eq.\eqref{2.115}. So, we have the following line element given by
\begin{equation}
ds^2= -c^2\left(1-\frac{2\alpha}{c^2}\right)dt^2 - \frac{4}{c}(\vec{A} \cdot d\vec{r})dt +\left(1+\frac{2\beta}{c^2}\right)dr^2.
\end{equation}
Now, the second term in \eqref{3.7} is zero as $\beta$ doesn't depend on time, and hence \eqref{3.7} becomes
\begin{equation}
\overset{3}{\Gamma_{i0}^j}=\frac{1}{c}\left[\frac{\partial A_i}{\partial x^j}-\frac{\partial A_j}{\partial x^i}\right].
\end{equation}
and the component \eqref{3.8} becomes
\begin{equation}
\overset{2}{\Gamma_{i0}^0}=\frac{1}{c^2}\frac{\partial \alpha}{\partial x^i},
\end{equation}
while, the component \eqref{3.9} becomes
\begin{equation}
\overset{2}{\Gamma_{ik}^j}=\frac{1}{c^2}\left[-\delta_{ij} \frac{\partial \beta}{\partial x^k}-\delta_{jk} \frac{\partial \beta}{\partial x^i}+\delta_{ik} \frac{\partial \beta}{\partial x^j}\right].
\end{equation}
Hence, the first term of \eqref{3.14} will yield the following 
\begin{equation}
\overset{3}{\Gamma_{i0}^j}S_j=\left[\frac{1}{c} \vec{S}\times(\vec{\nabla} \times \vec{A})\right]_i.
\end{equation}
Now if we look at the second term of \eqref{3.14}, it becomes
\begin{equation}
\overset{2}{\Gamma_{i0}^0}v^jS_j=\frac{1}{c^2}\frac{\partial \alpha}{\partial x^i}v^jS_j=\left[\frac{1}{c^2}(\vec{v}\cdot\vec{S})\vec{\nabla}\alpha\right]_i,
\end{equation}
while the third term of \eqref{3.14} can be written as
\begin{align}
\overset{2}{\Gamma_{ik}^j}v^kS_j &=\frac{1}{c^2}\left[-\delta_{ij} \frac{\partial \beta}{\partial x^k}v^kS_j-\delta_{jk} \frac{\partial \beta}{\partial x^i}v^kS_j+\delta_{ik} \frac{\partial \beta}{\partial x^j}v^kS_j\right]\\
&=\frac{1}{c^2}\left[-\frac{\partial \beta}{\partial x^k}v^kS_i-\frac{\partial \beta}{\partial x^i}v^kS_k+\frac{\partial \beta}{\partial x^j}v_iS_j\right]\\
&=\frac{1}{c^2}\left[-(\vec{v}\cdot\vec{S})\vec{\nabla}\beta-\vec{S}(\vec{v}\cdot\vec{\nabla}\beta)+\vec{v}(\vec{S}\cdot\vec{\nabla}\beta)\right].
\end{align}
As a result of the above calculations, \eqref{3.14} will become
\begin{equation}\label{3.24}
\frac{d\vec{S}}{dt}= \frac{1}{c} \vec{S}\times(\vec{\nabla} \times \vec{A})-\frac{1}{c^2}(\vec{v}\cdot\vec{S})\vec{\nabla}\alpha +\frac{1}{c^2}\left[-(\vec{v}\cdot\vec{S})\vec{\nabla}\beta-\vec{S}(\vec{v}\cdot\vec{\nabla}\beta)+\vec{v}(\vec{S}\cdot\vec{\nabla}\beta)\right].
\end{equation}
To solve \eqref{3.24}, we use the fact that parallel transport preserves the value of $S_{\mu}S^{\mu}$, so that we will have
\begin{equation}\label{3.25}
\frac{d}{dt}(g^{\mu\nu}S_\mu S_\nu)=0.
\end{equation}
The rate of change of $\vec{S}$ as seen from \eqref{3.14} is $\vec{S}$ times $\dfrac{\vec{v}^3}{r}$, so we want to keep the terms in $g_{\mu\nu}-\eta_{\mu\nu}$ whose rate of change is comparable as seen by a particle moving with velocity $\vec{v}$, i.e., those terms whose gradient is of order $\dfrac{\vec{v}^2}{r}$. Here, $g^{\mu\nu}$ may be replaced in equation \eqref{3.25} with $\eta^{\mu\nu}+h^{\mu\nu}$. Furthermore, $S_0^2$ is already of order $\vec{v}^2$ with respect to $\vec{S}^2$, so we need not keep $h^{00}$. So, finally, we will have
\begin{equation}
(\eta^{\mu\nu} +h^{\mu\nu})S_{\mu}S_{\nu}=constant,
\end{equation}
\begin{equation}
-S_0^2+\vec{S}^2+\frac{2\beta}{c^2}\vec{S}^2=constant.
\end{equation}
Using \eqref{3.12} in above equation, we get
\begin{equation}
\vec{S}^2+\frac{2\beta}{c^2}\vec{S}^2-(\vec{v}\cdot\vec{S})^2=constant.
\end{equation}
From now onward, we put $c=1$ for sake of convenience albeit we shall introduce it later for dimensional consistency.\\\\
As done in case of GR \cite{weinberg2014gravitation}, we shall start by introducing a new spin vector $\vec{\zeta}$, such that
\begin{equation}\label{3.29}
\vec{S}=(1-\beta)\vec{\zeta}+\frac{1}{2}\vec{v}(\vec{v}\cdot\vec{\zeta}).
\end{equation}
To the required order, we can invert equation \eqref{3.29} trivially by vector multiplication and use the properties of vectors to see that
\begin{equation}\label{49}
\vec{\zeta}=(1+\beta)\vec{S}-\frac{1}{2}\vec{v}(\vec{v}\cdot\vec{S}).
\end{equation}
Similarly, one can also check that ${\vec{\zeta}}^2= constant$. The rate of change of $\vec{\zeta}$ is given to order $\dfrac{\vec{v}^3}{r}\vec{S}$ by treating $\vec{S}$ as constant everywhere it appears with coefficients of order $\vec{v}^2$ which can then be written as
\begin{equation}
\frac{d\vec{\zeta}}{dt}=  \frac{d\vec{S}}{dt}+\vec{S}\left(\frac{\partial\beta}{\partial t}+\vec{v}\cdot\vec{\nabla}\beta\right)-\frac{1}{2}\frac{d\vec{v}}{dt}\left(\vec{v}\cdot\vec{S}\right) -\frac{1}{2}\vec{v}\left(\frac{d\vec{v}}{dt}\cdot\vec{S}\right)-\frac{1}{2}\vec{v}\left(\vec{v}\cdot\frac{d\vec{S}}{dt}\right).
\end{equation}
Since we are interested in determining this quantity to order $\dfrac{\vec{v}^3}{r}\vec{S}$, the last term is neglected as it is of higher order than required and the first component in the second term in the above equation is put to zero as $\beta$ doesn't depend on time. So, the above expression simplifies to
\begin{equation}
\frac{d\vec{\zeta}}{dt}=\frac{d\vec{S}}{dt}+\vec{S}(\vec{v}\cdot\vec{\nabla}\beta)-\frac{1}{2}\frac{d\vec{v}}{dt}\left(\vec{v}\cdot\vec{S}\right)-\frac{1}{2}\vec{v}\left(\frac{d\vec{v}}{dt}\cdot\vec{S}\right).
\end{equation}
Now, by setting $\dfrac{d\vec{v}}{dt}=-\vec{\nabla}\alpha$, the above expression yields
\begin{equation}
\frac{d\vec{\zeta}}{dt}=\frac{d\vec{S}}{dt}+\vec{S}(\vec{v}\cdot\vec{\nabla}\beta)+\frac{1}{2}\vec{\nabla}\alpha(\vec{v}\cdot\vec{S})+\frac{1}{2}\vec{v}(\vec{\nabla}\alpha\cdot\vec{S}).
\end{equation}
Using equation \eqref{3.24} in above equation, we get
\begin{equation}\label{3.36}
\frac{d\vec{\zeta}}{dt}= \vec{S}\times(\vec{\nabla} \times \vec{A})-\frac{1}{2}(\vec{v}\cdot\vec{S})\vec{\nabla}\alpha-(\vec{v}\cdot\vec{S})\vec{\nabla}\beta  +\vec{v}(\vec{S}\cdot\vec{\nabla}\beta)+\frac{1}{2}\vec{v}(\vec{\nabla}\alpha\cdot\vec{S}).
\end{equation}
At this point, one can check whether the above calculation is correct by setting $\vec{\nabla}\alpha$ and $\vec{\nabla}\beta$ as $\vec{\nabla}\Phi$ where $\Phi=\dfrac{GM}{r}$ and confirming it with the GR result as given in \cite{weinberg2014gravitation}.\\\\
Now, from the definitions of $\alpha$ and $\beta$ already introduced, \eqref{2.110} and \eqref{2.111}, let us rewrite this quantities as
\begin{equation}
\alpha=\Phi+\Phi F,
\end{equation}
and
\begin{equation}
\beta=\Phi-\Phi F,
\end{equation}
with $F=\dfrac{\exp{(-\Upsilon r)}}{3}$. This implies that $\beta$ can be expressed in terms of $\alpha$ trivially as
\begin{equation}
\beta=2\Phi-\alpha,
\end{equation}
and, so, after some calculations and replacing the value of $\alpha$ again, equation \eqref{3.36} can be rewritten as
\begin{multline}
\frac{d\vec{\zeta}}{dt}=\vec{S}\times(\vec{\nabla} \times \vec{A})-\frac{3}{2}(\vec{v}\cdot\vec{S})\vec{\nabla}\Phi+\frac{3}{2}\vec{v}(\vec{S}\cdot\vec{\nabla}\Phi)  +\frac{1}{2}(\vec{v}\cdot\vec{S})\vec{\nabla}(\Phi F)-\frac{1}{2}\vec{v}(\vec{S}\cdot\vec{\nabla}(\Phi F)).
\end{multline}
To order $\dfrac{\vec{v}^3}{r}\vec{\zeta}$, we can just replace $\vec{S}$ with $\vec{\zeta}$ given that by doing so we only miss higher order terms. So, after using some vector identities, we end up with the following expression for the rate of change of the $\vec{\zeta}$
\begin{equation}\label{3.49}
\frac{d\vec{\zeta}}{dt}= {\vec{\zeta}}\times(\vec{\nabla} \times \vec{A})+\frac{3}{2}\vec{\zeta}\times(\vec{v}\times\vec{\nabla}\Phi)  +\frac{1}{2}\left[\vec{\zeta}\times(\vec{\nabla}(\Phi F)\times\vec{v} )\right],
\end{equation}
which can be expressed concisely by introducing the following quantities  
\begin{equation}\label{3.50}
\vec{\Omega_{GR}}\equiv-\vec{\nabla}\times\vec{A}-\frac{3}{2}\vec{v}\times\vec{\nabla}\Phi,
\end{equation}
which corresponds to the angular frequency of precession for a gyroscope in GR \cite{weinberg2014gravitation}, and 
\begin{equation}\label{3.52}
\vec{\Omega_{f(R)}}\equiv \vec{\Omega_{GR}}-\frac{1}{2}\vec{\nabla}(\Phi F)\times\vec{v}.
\end{equation}
which is the Euler's rotation in the absence of torque and where $\vec{\Omega_{f(R)}}$ is the $f(R)$ corrected angular frequency.
Indeed, doing so equation \eqref{3.49} takes the simple form
\begin{equation}
\frac{d\vec{\zeta}}{dt}=\vec{\Omega_{f(R)}}\times\vec{\zeta},
\end{equation}

In our calculation, we have the following quantities as discussed in the section \ref{1}
\begin{equation}
\Phi=\frac{GM}{r},
\end{equation}
\begin{equation}\label{3.54}
\vec{A}=\frac{G}{r^3}\left(\vec{r}\times\vec{J}\right).
\end{equation}
After executing a series of vector manipulations,\eqref{3.52} gives us (See appendix \ref{rudra})
\begin{equation}\label{3.69}
\vec{\Omega_{f(R)}}=  3G\frac{\vec{r}(\vec{r}\cdot\vec{J})}{r^5}-G\frac{\vec{J}}{r^3}-\frac{3GM}{2r^3}(\vec{r}\times\vec{v}) +\frac{1}{2}\left[\frac{GM}{r^2}\left(\frac{1}{r}\frac{\exp(-\Upsilon r)}{3}+\frac{\Upsilon}{3}\exp(-\Upsilon r)\right)\right]\left[\vec{r}\times\vec{v}\right].
\end{equation}
The first three terms are exactly the same as in GR case \cite{weinberg2014gravitation}\footnote{Notice that differently from \cite{weinberg2014gravitation}, we have a minus sign before the third term in \eqref{3.69} because we considered the potential with an overall plus sign}. So, we can now write it as
\begin{equation}
\vec{\Omega_{f(R)}}=\vec{\Omega_{GR}}+\frac{1}{2}\left[\frac{GM}{r^2}\frac{\exp(-\Upsilon r)}{3}\left(\frac{1}{r}+\Upsilon\right)\right]\left[\vec{r}\times\vec{v}\right].
\end{equation}
This is the expression for the  metric $f(R)$ corrected gyroscopic frequency when $f(R)$ is analytic. The first two terms in \eqref{3.69} represents an interaction between the spin orbital angular momenta of the Earth and the gyroscope and are responsible for the so called Lense--Thirring Precession. The last two terms which depends only on the mass of the Earth and not on the spin make up the so called Geodetic Precession.
\subsection{\textbf{Discussion on the Geodetic precession frequency}}
\begin{itemize}
\item $f(R)$ gives a contribution to the geodetic precession, i.e., to the third term in the expression \eqref{3.69}.\\
\item However, $f(R)$ doesn't alter the Lense--Thirring precession (first two terms in the expression \eqref{3.69} which is also a consequence of the result that taking an analytical expansion of $f(R)$ gives us the exact similar linear equations to GR. \\
\item $f(R)$ increases the total precession frequency but decreases the absolute value of geodetic frequency since we have\\\\ $\vec{\Omega}_{geodetic}=\frac{3GM}{2r^3}(\vec{r}\times\vec{v})-\frac{1}{2}\left[\frac{GM}{r^2}\left(\frac{1}{r}\frac{\exp(-\Upsilon r)}{3}+\frac{\Upsilon}{3}\exp(-\Upsilon r)\right)\right]\left[\vec{r}\times\vec{v}\right]$.\\\\ But the correction is quite small if large distances are taken in consideration. 
\end{itemize}
\section{Constraints on \textbf{$\Upsilon$} from Gravity Probe--B (GP--B)}
If for simplicity, we take the gyroscope orbit to be circular with radius $r$ and unit vector $\vec{\hat{k}}$ to be the normal to the orbit, we have the following expression for velocity
\begin{equation}
\vec{v}=-\left(\frac{GM}{r^3}\right)^{\frac{1}{2}}(\vec{r}\times\vec{\hat{k}}).
\end{equation}
But the expression for $\vec{\Omega}_{geodetic}$ as written in the third point of the above discussion is not enough for the experimental verification as we have derived the results for a spherical Earth whereas in practice, Earth is not spherical but oblate. Hence, we have to take in consideration the oblateness of Earth. For that purpose we will make the following consideration:\\\\
Since the $f(R)$ corrections are relevant at small $\Upsilon$ and small distances, the third term in $\vec{\Omega}_{geodetic}$  will be negligible as compared to the first two terms. So, for the following calculations, we will consider only the first two terms.\\\\
As a result, we have the following angular frequency
\begin{equation}\label{3.75}
\begin{split}
\vec{\Omega}_{geodetic}&=\left[\frac{3}{2c^2}-\frac{1}{6c^2}\exp(-\Upsilon r)\right]\frac{GM}{r^3}(\vec{r}\times\vec{v})\\
&=\left[\frac{3}{2c^2}-\frac{1}{6c^2}\exp(-\Upsilon r)\right](\vec{g}\times\vec{v}).
\end{split}
\end{equation}
Where, we have defined $\vec{g}\equiv \dfrac{GM}{r^3}\vec{r}$, the GR gravitational acceleration at the location of the gyroscope and we have introduced the square of the velocity of light, $c^2$, for dimensional consideration.\\\\
The total geodetic precession (GR term plus the oblateness correction) was calculated in \cite{wilkins1970general} and \cite{barker1970derivation} using rather elaborate analytical techniques. Here, we use the convenient derivation and terminologies given in \cite{breakwell1988stanford} to extend it to the $f(R)$ correction\\\\
To calculate the $f(R)$ corrected geodetic term along with the oblateness correction, we compute, $\vec{g}\times\vec{v}$, for an actual orbit around the oblate Earth (See appendix \ref{shiva}).\\\\
While doing so, we neglect the second order terms in the Earth's quadrupole moment, $J_2$, and the mean eccentricity, $e$, and go through the calculations given in \cite{breakwell1988stanford} but using our expression for $\Omega_{geodetic}$ \eqref{3.75}. Since the satellite was inserted in the polar orbit, we consider only the polar orbit result for the geodetic angular frequency. Hence, we arrive at the following expression after taking the average per orbit \ref{shiva}.
\begin{equation}\label{3.88}
\left\langle\vec{\Omega}\right\rangle_{geodetic}|_{polar}= \left[\frac{3}{2c^2}-\frac{1}{6c^2}\exp(-\Upsilon r)\right] \frac{(GM)^{\frac{3}{2}}}{\bar{r}^{\frac{5}{2}}}\left[1-\frac{9}{8}J_2\left(\frac{R_e}{\bar{r}}\right)^2\right]\vec{\hat{k}}.
\end{equation}
For further calculations, we need the knowledge of the actual Cartesian inertial frame used by GP--B for data reduction \cite{silbergleit2015gravity}.
Let us write $x = x_1, y = x_2, z = x_3$ for the Cartesian coordinates of the inertial frame JE2000, with the unit vectors $\hat{x} = \hat{x}_1, \hat{y} = \hat{x}_2, \hat{z} = \hat{x}_3$ along the corresponding axes.
 It is natural to set one axis in the direction of the guide star (GS) with unit vector $\vec{\hat{e}_{\rm{gs}}}$. 
 The z axis of the JE2000 frame was exactly parallel to the Earth rotation axis on noon GMT, 1 January 2000, and stayed within a few arc-seconds (1as$= 4.848 \times 10^{-6}$ rad) throughout the entire GPÐB flight in 2004-2005.

 The ideal GP--B polar orbit would contain both $\vec{\hat{e}_{\rm{gs}}}$ and $\vec{\hat{z}}$ \cite{silbergleit2015gravity}, hence a good choice for the second unit vector of the frame under construction is $\vec{\hat{e}_{\rm{we}}}$. 
  \begin{equation}
  \vec{\hat{e}_{\rm{we}}}=\frac{\vec{\hat{e}_{\rm{gs}}}\times\vec{\hat{z}}}{|\vec{\hat{e}_{\rm{gs}}}\times\vec{\hat{z}}|}.
  \end{equation}
 The index WE stands for West--East direction perpendicular to the ideal orbit plane, in which the gyroscope drifts due to the relativistic Lense--Thirring effect. The third axis is defined in the usual way:
\begin{equation}\label{3.91}
\vec{\hat{e}_{\rm{ns}}}=\vec{\hat{e}_{\rm{we}}}\times\vec{\hat{e}_{\rm{gs}}}.
\end{equation}
The unit vector $\vec{\hat{e}_{\rm{ns}}}$ lies in the ideal orbit plane and is orthogonal to $\vec{\hat{e}_{\rm{gs}}}$, the geodetic relativistic drift goes in the NS (North--South) direction.\\\\
The $\vec{\hat{k}}$ in the equation \eqref{3.88} is indeed the $\vec{\hat{e}_{\rm{we}}}$ direction for the GP--B coordinates. \\\\
As such, for the precession rate $\vec{R}\equiv\frac{d\vec{\zeta}}{dt}=\Omega\times\vec{\zeta}$, we get
\begin{equation}
\begin{aligned}
\vec{R}_{geodetic}=& \left[\frac{3}{2c^2}-\frac{1}{6c^2}\exp(-\Upsilon \bar{r})\right]\frac{(GM)^{\frac{3}{2}}}{\bar{r}^{\frac{5}{2}}}\left[1-\frac{9}{8}J_2\left(\frac{R_e}{\bar{r}}\right)^2\right][\vec{\hat{e}_{\rm{we}}}\times\vec{\hat{e}_{\rm{gs}}}]\\
&=\left[\frac{3}{2c^2}-\frac{1}{6c^2}\exp(-\Upsilon \bar{r})\right] \frac{(GM)^{\frac{3}{2}}}{\bar{r}^{\frac{5}{2}}}\left[1-\frac{9}{8}J_2\left(\frac{R_e}{\bar{r}}\right)^2\right]\vec{\hat{e}_{\rm{ns}}}.
\end{aligned}
\end{equation}
Where,
\begin{equation}
\vec{R}_{GR}=\frac{3}{2c^2}\frac{(GM)^{\frac{3}{2}}}{\bar{r}^{\frac{5}{2}}}\left[1-\frac{9}{8}J_2\left(\frac{R_e}{\bar{r}}\right)^2\right]\vec{\hat{e}_{\rm{ns}}},
\end{equation}
was reported to be $6606.1$ mas/yr (milli arc second per year) with $1\sigma$ uncertainties \cite{silbergleit2015gravity} ($J_2\approx1.083\times10^{-3}$)\\\\
The final joint result for all the gyroscopes indicate that $\vec{R_{NS,obs}}=6601.8\pm18.3$ mas/yr \cite{overduin2013constraints}\\\\
So, the NS components of the relativistic drift may deviate from predictions of GR by at most 
\begin{align}
\Delta \vec{R_{NS}} & <|\vec{R_{GR}}-\vec{R_{NS,obs}}|=\SI{22.6}{\frac{mas}{yr}},
\end{align}
and we have the following constraint
\begin{equation}
\frac{1}{6c^2}\exp(-\Upsilon \bar{r})\frac{(GM)^{\frac{3}{2}}}{\bar{r}^{\frac{5}{2}}}\left[1-\frac{9}{8}J_2\left(\frac{R_e}{\bar{r}}\right)^2\right]<22.6.
\end{equation}
After using standard values for the quantities and $\bar{r}=7018$ $\si{km}$, we get $1.3048\times10^{-13} \exp(-\Upsilon \bar{r})$  $\si{\radian\per\second}$. We have to convert it into mas/year. So, after doing the conversion, we get $735 \exp(-\Upsilon \bar{r})$ mas/yr, and finally obtain
\begin{equation}
735\exp(-\Upsilon \bar{r})<22.6,
\end{equation}
which implies $\Upsilon > 0.5\times10^{-6}$ $\si{m}^{-1}$ and hence $|a_2|<1.33\times10^{12}$ $ \si{m^2}$.
\subsection{\textbf{Discussion on the constraints of $\Upsilon$}}
In \cite{berry2011linearized}, $a_2$ was computed for various test. Considering the phase of a gravitational waveform, estimated deviations from general relativity could be measurable for an extreme--mass--ratio inspiral about a $10^6 M_{\odot}$ black hole if $|a_2|\gtrsim 10^{17} \si{m^2}$, assuming that the weak--field metric of the black hole coincides with that of a point mass which seems to rule out GR in the particular regime. While the planetary precession gave a bound of $|a_2|\lesssim 1.2\times10^{18} \si{m^2}$ \cite{berry2011linearized},
and the strongest constraint was placed by E\"ot--Wash experiment \cite{kapner2007tests,hoyle2004submillimeter}, a laboratory experiment giving a constraint $|a_2|\lesssim 2\times10^{-9} \si{m^2}$. A similar bound is quoted in \cite{naf20101}.
Our calculation for the geodetic precession, gives a bound intermediate between the planetary precession and E\"ot--Wash experiment, i.e.~$|a_2|<1.33\times10^{12} \si{m^2}$.\\\\
Indeed, our satellite--scale observation is much weaker than the laboratory bounds. However, it is still of investigative interest as it probes gravity at a different scale and in a different environment altogether. Moreover, we cannot assume $f(R)$--gravity to be universal, as one cannot exclude that it may be different in different regions of space or it may vary with the energy scales. The limits on $a_2$ from GP--B depends on several parameters pertaining to the satellite like the orbital radius. However, if the laboratory bound is indeed universal, observation of a deviation would mean that GR failed and suggest that $a_2$ would have to vary with the environment. Along with that, in Quantum field theory, all couplings run with energy \cite{peskin2018introduction}. Hence, it is possible that $a_2$ depends on the energy scales of the environment and therefore on the position in space--time. \\\\
A chameleon mechanism could help us in explaining the variation \cite{berry2011linearized}, where f(R)--gravity is modified in the presence of matter \cite{khoury2004chameleon,khoury2004chameleon2,brax2004detecting} and the metric $f(R)$ has a non--linear effect which arises from the large departure of the Ricci scalar from the background value \cite{de2010f}. Also, the mass of the effective scalar degree of freedom depends on the density of the environment \cite{faulkner2007constraining,li2007cosmology}. On Earth, we have high density and hence a high $\Upsilon$ or frequency of the scalar mode which suppresses the deviations from GR, while at the scale of GP--B, we have a relatively low density and hence a small $\Upsilon$. \\\\
Also, we make a note here that, metric $f(R)$ with $f$ analytic does not provide us with an improvement on the Lense--Thirring effect as it can be seen from the first two terms of the gyroscopic precession \eqref{3.69}. However, we will see in the following section, that an improvement can indeed be expected by considering a Brans--Dicke theory with a potential.

\section{\textbf{Generalisation to Brans--Dicke theories with a potential}}\label{durga}
It is now natural to ask what would be the gyroscopic precession frequency in a more generalised setting given that metric $f(R)$ is a special case of the Brans--Dicke theories \cite{sotiriou2010f}.
The Brans--Dicke action with an arbitrary potential in Jordan frame is given as 
\begin{equation}
\mathcal{S}=\int d^4x\sqrt{-g}\left\{\frac{1}{16\pi}\left[\phi R-\frac{\omega}{\phi}g^{\mu\nu}\partial_{\mu}\phi\partial_{\nu}\phi-V(\phi)\right]+ \mathcal{L}_{matter}\right\},
\end{equation}
where the scalar field, $\phi$, is coupled by a dimensionless constant called the Brans--Dicke parameter, $\omega$, and $V(\phi)$ is an arbitrary potential. The scalar field $\phi$ does not have the canonical dimension one and instead has dimension two like that of the Newton's constant.\\\\
The field equations are 
\begin{equation}
G_{\mu\nu}= \frac{8\pi}{\phi}T_{\mu\nu}+\frac{\omega}{\phi^2}\left(\nabla_{\mu}\phi\nabla_{\nu}\phi-\frac{1}{2}g_{\mu\nu}\nabla^{\alpha}\phi\nabla_{\alpha}\phi\right) + \frac{1}{\phi}\left(\nabla_{\mu}\phi\nabla_{\nu}\phi-g_{\mu\nu}\Box_g\phi\right)-\frac{V(\phi)}{2\phi}g_{\mu\nu},
\end{equation}
and 
\begin{equation}
\Box_g\phi=\frac{1}{2\omega+3}\left(8\pi T+\phi\frac{dV(\phi)}{d\phi}-2V(\phi)\right),
\end{equation}
where $T$ is the trace of the matter energy momentum tensor, $T_{\mu\nu}$, and $\Box_g$ is the d'Alembertian operator with respect to the Jordan metric. Let us investigate how the weak field equations look like \cite{ozer2018linearized,dass_liberati2}. As such, we consider the following expansion
\begin{align}
g_{\mu\nu} &=\eta_{\mu\nu}+h_{\mu\nu}, & \phi=\phi_0+\xi,
\end{align}
where $\phi_0$ is a constant value of the scalar field and $\xi$ is the small perturbation to the scalar field, while rest of the symbols have their usual meaning. A new tensor can be defined as the following \cite{will1993theory}
\begin{equation}\label{2.152}
\theta_{\mu\nu}=h_{\mu\nu}-\frac{1}{2}\eta_{\mu\nu}h-\eta_{\mu\nu}\frac{\xi}{\phi_0},
\end{equation}
and for which the Brans--Dicke gauge must be true
\begin{equation}
\nabla_{\nu}\theta^{\mu\nu}=0.
\end{equation}
The weak field equations up to second order are
\begin{equation}\label{2.154}
\Box_{\eta}\theta_{\mu\nu}=-\frac{16\pi}{\phi_0}(T_{\mu\nu}+\tau_{\mu\nu})+\frac{V_{\mathrm{lin}}}{\phi_0}g_{\mu\nu},
\end{equation}
and
\begin{equation}\label{2.155}
\Box_{\eta}\xi=16\pi S.
\end{equation}
Here $\Box_{\eta}=\eta^{\mu\nu}\partial_{\mu}\partial_{\nu}$ is d'Alembertian of the flat spacetime and other symbols have their usual meaning, while the term $S$ is given by
\begin{equation}
S=\frac{1}{4\omega+6}\left[T\left(1-\frac{\theta}{2}-\frac{\xi}{\phi_0}\right)+\frac{1}{8\pi}\left(\phi\frac{dV}{d\phi}-2V\right)_{\mathrm{lin}}\right] +\frac{1}{16\pi}\left(\theta^{\mu\nu}\partial_{\mu}\partial_{\nu}\xi+\frac{\partial_{\nu}\xi\partial^{\nu}\xi}{\phi_0}\right).
\end{equation}
Here the subtext {$\mathrm{lin}$ means that the terms must be properly linearised.\\\\
In this derivation the relation between flat and curved spacetime d'Alembertians is used
\begin{equation}
\Box_{g}=\left(1+\frac{\theta}{2}+\frac{\xi}{\phi_0}\right)\Box_{\eta}-\theta^{\mu\nu}\partial_{\mu}\partial_{\nu}\xi-\frac{\partial_{\nu}\xi\partial^{\nu}\xi}{\phi_0} +\mathcal{O}{({x_{i}}^2)}.
\end{equation}
The arbitrary potential $V$ is assumed to be a well behaved function and it is Taylor expandable around a constant value $\phi=\phi_0$ such that \cite{ozer2018linearized}
\begin{equation}
V(\phi)=V(\phi_0)+\frac{dV(\phi_0)}{d\phi}\xi+\frac{1}{2}\frac{d^2V(\phi_0)}{d\phi^2}\xi^2+...
\end{equation}
Here $\phi_0$ is the expected minimum of the potential and hence the term $\dfrac{dV(\phi_0)}{d\phi}$ vanishes. Hence, the relevant terms in the linearised equation can be written as
\begin{align}
V(\phi)g_{\mu\nu} &\approx V(\phi_0)\eta_{\mu\nu} \nonumber,\\ 
 \left(\phi\frac{dV}{d\phi}-2V\right) & \approx\phi_0 \frac{d^2V(\phi_0)}{d\phi^2}\xi-2V(\phi_0),
\end{align}
and hence the field equations \eqref{2.154} and \eqref{2.155} become
\begin{equation}\label{2.160}
\Box_{\eta}\theta_{\mu\nu}=-\frac{16\pi}{\phi_0}T_{\mu\nu}+\frac{V_0}{\phi_0}\eta_{\mu\nu},
\end{equation}
and
\begin{equation}\label{2.161}
(\Box_{\eta}-m_s^2)\xi=\frac{8\pi T}{2\omega+3}-\frac{2V_0}{2\omega+3},
\end{equation}
where 
\begin{align}\label{2.162}
V_0 &\equiv V(\phi_0), & m_s^2 \equiv \frac{\phi_0}{2\omega+3}\frac{d^2V(\phi_0)}{d\phi^2}>0.
\end{align}
A particle located at $\bar{r}=0$ is considered, where $\bar{r}^2=\bar{x}^2+\bar{y}^2+\bar{z}^2$ and $T_{\mu\nu}=M\delta(\bar{r})$.
The solution of the scalar field equation \eqref{2.161} is given by
\begin{equation}
\xi(\bar{r})=\frac{2M}{(2\omega+3)}\frac{\exp(-m_s \bar{r})}{\bar{r}}-\frac{V_0}{3(2\omega+3)}\bar{r}^2.
\end{equation}
The solution to \eqref{2.160} are 
\begin{align}
\theta_{00} &=-\frac{4M}{\phi_0}\frac{1}{\bar{r}}+\frac{V_0}{6\phi_0}\bar{r}^2, & \theta_{xx} &=-\frac{V_0}{4\phi_0}(y^2+z^2),\\
\theta_{xx} &=-\frac{V_0}{4\phi_0}(x^2+z^2), & \theta_{xx} &=-\frac{V_0}{4\phi_0}(x^2+y^2).
\end{align}
The trace $\theta$ is given by
\begin{equation}
\theta=-\frac{4M}{\phi_0\bar{r}}+\frac{2V_0}{3\phi_0}\bar{r}^2,
\end{equation}
and from the inverse of \eqref{2.152}, we get
\begin{align}
h_{00} &=-\left[\frac{2M}{\phi_0\bar{r}}+\frac{V_0}{6\phi_0}\bar{r}^2+\frac{\xi}{\phi_0}\right],\\
h_{ij} &=-\left[\frac{2M}{\phi_0\bar{r}}-\frac{V_0}{12\phi_0}(\bar{r}^2+3x_i^2)-\frac{\xi}{\phi_0}\right]\delta_{ij}.
\end{align}
To express the solution in isotropic coordinates, the following transformation is employed \cite{bernabeu2010cosmological}
\begin{equation}
\bar{x}^{i}=x^{i}+\frac{V_0}{24\phi_0}{x^{i}}^3.
\end{equation}
We then get
\begin{align}
h_{00} &=-\left(\frac{2M}{\phi_0 r}+\frac{V_0}{6}r^2+\frac{\xi}{\phi_0}\right),\\
h_{ij} &=-\left(\frac{2M}{\phi_0 r}-\frac{V_0}{12}r^2-\frac{\xi}{\phi_0}\right)\delta_{ij},\\
\xi &=\frac{2M}{(2\omega+3)}\frac{\exp(-m_s r)}{r}-\frac{V_0}{3(2\omega+3)}r^2.
\end{align}
The full metric components are given by
\begin{equation}
g_{00}=  1-\frac{2M}{\phi_0 r}\left(1+\frac{\exp(-m_s r)}{2\omega+3}\right)- \frac{V_0 r^2}{6\phi_0}\left(1-\frac{2}{2\omega+3}\right),
\end{equation}
\begin{equation}
g_{ij}=  -\left[1+\frac{2M}{\phi_0 r}\left(1-\frac{\exp(-m_s r)}{2\omega+3}\right) -\frac{V_0 r^2}{12\phi_0}\left(1-\frac{4}{2\omega+3}\right)\right]\delta_{ij},
\end{equation}
\begin{equation}
\phi= \phi_0\left(1+\frac{2M}{(2\omega+3)}\frac{\exp(-m_s r)}{\phi_0 r}-\frac{V_0}{3\phi_0(2\omega+3)}r^2\right).
\end{equation}
Now to compute the precession frequency, we follow the steps previously done for metric $f(R)$. And, hence, without going into explicit calculations, we can directly recall \eqref{3.36} for the rate of change of the defined spin vector $\zeta$ which is given by
\begin{equation*}
\frac{d\vec{\zeta}}{dt}=  \vec{S}\times(\vec{\nabla} \times \vec{A})-\frac{1}{2}(\vec{v}\cdot\vec{S})\vec{\nabla}\alpha-(\vec{v}\cdot\vec{S})\vec{\nabla}\beta +\vec{v}(\vec{S}\cdot\vec{\nabla}\beta)+\frac{1}{2}\vec{v}(\vec{\nabla}\alpha\cdot\vec{S}),
\end{equation*}
where now the definitions of $\alpha$ and $\beta$ will change in accordance with the line elements. In this case, those are given by the following two expressions
\begin{equation}
\alpha\equiv\frac{M}{\phi_0 r }\left(1+\frac{\exp(-m_s r)}{2\omega+3}\right)+\frac{V_0 r^2}{12\phi_0}\left(1-\frac{2}{2\omega+3}\right),
\end{equation}
and
\begin{equation}
\beta\equiv\frac{M}{\phi_0 r}\left(1-\frac{\exp(-m_s r)}{2\omega+3}\right)-\frac{V_0 r^2}{24\phi_0}\left(1-\frac{4}{2\omega+3}\right).\\
\end{equation}
Let us now define the following quantities, just like we did in the previous section of metric $f(R)$ 
\begin{align}
\alpha &=\kappa+\kappa\mathcal{F}+\mathcal{N},\\
\beta &=\kappa-\kappa\mathcal{F}-\mathcal{M}.
\end{align}
where, we have defined
\begin{align}\label{3.103}
\kappa &\equiv\dfrac{M}{\phi_0 r},\nonumber \\
\mathcal{F} &\equiv \dfrac{\exp(-m_s r)}{2\omega+3}\nonumber,\\
\mathcal{N}&\equiv\dfrac{V_0 r^2}{12\phi_0}\left(1-\frac{2}{2\omega+3}\right)\nonumber,\\
\mathcal{M}&\equiv\frac{V_0 r^2}{24\phi_0}\left(1-\frac{4}{2\omega+3}\right).
\end{align}
Using the above notations, we can now write the rate of change of spin $\zeta$ as
\begin{equation}
\frac{d\vec{\zeta}}{dt}={\vec{\zeta}}\times(\vec{\nabla} \times \vec{A})+\frac{3}{2}\vec{\zeta}\times(\vec{v}\times\vec{\nabla}\kappa)+\frac{1}{2}\left[\vec{\zeta}\times(\vec{\nabla}(\kappa \mathcal{F})\times\vec{v} )\right] + \vec{\zeta}\times(\vec{v}\times\vec{\nabla}\mathcal{N})-\vec{\zeta}\times(\vec{v}\times\vec{\nabla}\mathcal{M}),
\end{equation}
which can then be conveniently expressed as 
\begin{equation}
\frac{d\vec{\zeta}}{dt}=\vec{\Omega_{BD}}\times\vec{\zeta},
\end{equation}
which is the Euler's rigid body rotation equation without torque and where $\vec{\Omega_{BD}}$ is the gyroscopic frequency in the Brans--Dicke theory 
\begin{equation}\label{3.109}
\vec{\Omega_{BD}}= -\vec{\nabla}\times\vec{A}-\frac{3}{2}\vec{v}\times\vec{\nabla}\kappa-\frac{1}{2}\vec{\nabla}(\kappa\mathcal{F})\times\vec{v}-(\vec{v}\times\vec{\nabla}\mathcal{N})+(\vec{v}\times\vec{\nabla}\mathcal{M}).
\end{equation}
To compute the above expression, we have to first calculate the gradient of the quantities defined in \eqref{3.103}
\begin{align}
\vec{\nabla}(\kappa\mathcal{F}) &=-\frac{M}{\phi_0 r^3}\frac{\exp(-m_s r)}{2\omega+3}\vec{r}-\frac{M}{\phi_0 r^2}m_s\frac{\exp(-m_s r)}{2\omega+3}\vec{r},\\
\vec{\nabla}\mathcal{N} &=\frac{V_0}{6\phi_0}\left(1-\frac{2}{2\omega+3}\right)\vec{r},\\
\vec{\nabla}\mathcal{M} &=\frac{V_0}{12\phi_0}\left(1-\frac{4}{2\omega+3}\right)\vec{r}.
\end{align}
The first term of \eqref{3.109} can be computed by replacing $G$ in \eqref{3.54} by $\dfrac{1}{\phi_0}$ such that
\begin{equation}
\vec{A}=\frac{1}{\phi_0 r^3}\left(\vec{r}\times\vec{J}\right).
\end{equation}
Following the steps previously done for the case of metric $f(R)$, where $f(R)$ is analytic function of the Ricci scalar, we finally get for the gyroscopic precession in the context of Brans-Dicke
\begin{equation}\label{3.110}
\begin{aligned}
\vec{\Omega_{BD}}=&3\frac{\vec{r}(\vec{r}\cdot\vec{J})}{\phi_0 r^5}-\frac{\vec{J}}{\phi_0r^3}-\frac{3}{2}\frac{M}{\phi_0 r^3}(\vec{r}\times\vec{v})+ \left[\frac{M}{\phi_0 r^3}\frac{\exp(-m_s r)}{2\omega+3}\vec{r}+\frac{M}{\phi_0 r^2}m_s\frac{\exp(-m_s r)}{2\omega+3}\vec{r}\right](\vec{r}\times\vec{v})\\&+\frac{V_0}{6\phi_0}\left(1-\frac{2}{2\omega+3}\right)(\vec{r}\times\vec{v})-\frac{V_0}{12\phi_0}\left(1-\frac{4}{2\omega+3}\right)(\vec{r}\times\vec{v}).
\end{aligned}
\end{equation}

Let us stress that this formula encompass a wider class of theories beyond the $f(R)$ ones. 
Nonetheless it can be used to double check our previous result Eq.~\eqref{3.69} by using the well known correspondence between $f(R)$ theories and a specific class of BD theories with a potential~\cite{sotiriou2010f}. We did so in Appendix \ref{app:2} and \ref{laxmi} where we show explicitly the consistency of the obtained results.\\\\
Remarkably, it can be easily seen from Eq.~\eqref{3.110} that Palatini $f(R)$, which corresponds to a class of BD theories with coupling parameter $\omega=-\dfrac{3}{2}$ \cite{sotiriou2010f} corresponds to a singular point, implying that the gyroscope precession in the case of Palatini $f(R)$ is at best ill defined or it cannot be derived as a limit from the BD result. As such we think it deserves further investigation.
\section{Discussion and Outlook}
In the work, we have presented the derivation of the gyroscopic precession frequency in the context of metric $f(R)$ theory. Since we had to match our results with the data of Gravity Probe--B mission, which is a precision experiment, we had to take into account the oblateness of Earth in our expression for the Geodetic precession frequency. As such, a concise derivation of the same was presented taking in consideration the quadrupole moment of Earth's potential. Adapting our derivation to the GP--B coordinate system, we could derive the constraint, $|a_2|<1.33\times 10^{12}$ $\si{m^2}$.\\\\
This is promising when compared to the astrophysical bounds so far provided by massive Black Holes and solar--system tests \cite{berry2011linearized} and at the same time, it complementary to the  the tighter bounds provided by laboratory tests like the E\"ot--Wash experiment \cite{berry2011linearized,naf20101}. Indeed, albeit our constraint is weaker than the lab one, still it is derived in a completely different regime of scales. 
Moreover, a chameleon mechanism could lend further justification to pursue the large scale constraints in parallel to the lab ones, leading to different values for $a_2$ depending on the density of the environment \cite{khoury2004chameleon,khoury2004chameleon2,brax2004detecting,faulkner2007constraining,li2007cosmology}.\\\\
We concluded the paper by generalising our calculation for Gyroscopic Precession to the wider class of Brans--Dicke theories with a potential and then subsequently verified its f(R) and GR limits.\\\\
Remarkably, this showed a potential issue with Palatini $f(R)$ which appears to be a singular limit for the Brans--Dicke formula of the gyroscope precession frequency. We think that this definitely deserve further investigation as it may signal an unphysical feature of Palatini $f(R)$ similar to the one identified e.g.~in \cite{sotiriou2010f}.\\\\
Let us stress, that Scalar--tensor theories and consequently generalised Brans--Dicke theories, prima facie appears to be far more interesting for phenonomenological tests  than $f(R)$ theories as they incorporate a wider range of corrections. In particular, a further extension of this work could be done by considering a generalised Scalar--Tensor theory such as Horndeski gravity (of course keeping in mind the recent constraints on the theory \cite{lombriser2016breaking,lombriser2017challenges,sakstein2017implications}and \cite{baker2017strong,creminelli2017dark} where the scalar field is responsible for dark energy).  Future work could also focus on finding constraints for $\Upsilon$ parameter from other missions, for example, \textit{Gaia} (spacecraft) \cite{brown2016gaia}, via measurements of the gravitational lensing. We hope that the present work will stimulate further investigations along these lines.


\begin{acknowledgements}
AD would like to thank Alessio Baldazzi for useful discussions. This work was done in SISSA, Italy as part of master's thesis by AD. 
\end{acknowledgements}

\section{Appendix}
\subsection{\textbf{Quadropole correction to geodetic frequency}}\label{shiva}
The exact derivation given in \cite{breakwell1988stanford} is followed but with our expression for gyroscope frequency in metric $f(R)$. We neglect the second order terms in the Earth's quadrupole moment, $J_2$, and the mean eccentricity, $e$. Then, a near circular orbit around the earth can be described by four equations:\\\\
The mean position describing a circle of radius $\bar{r}$ in a precessing plane with constant inclination $i$ (inclination with respect to the equatorial plane of earth), the circle being described at a constant rate \cite{breakwell1988stanford} is given by
\begin{equation}
\dot{\theta}=\sqrt{\frac{GM}{\bar{r}^3}}\left\{1+J_2\left(\frac{R_e}{\bar{r}}\right)\left[\frac{9}{4}-\frac{21}{8}\sin^2 i\right]\right\},
\end{equation}
while the precession rate about the North Pole is 
\begin{equation}
\dot{\lambda_A}=-\frac{3}{2}J_2\left(\frac{R_e}{\bar{r}}\right)^2\dot{\theta}\cos i,
\end{equation}
where $M$ and $R_e$ are the mass and the mean equatorial radius of the Earth respectively.\\\\
The actual position of the satellite is displaced from the mean position by $\delta r$ and $\delta \theta$ in the precessing plane, as follows \cite{breakwell1988stanford}
\begin{equation}\label{3.78}
\delta r=\bar{r}\left\{\frac{1}{4}J_2\left(\frac{R_e}{\bar{r}}\right)^2\sin^2 i \cos 2\theta-e \cos(\theta-\theta_p)\right\},
\end{equation}
\begin{equation}\label{3.79}
\delta\theta=\frac{1}{8}\left(\frac{R_e}{\bar{r}}\right)^2\sin^2 i \sin 2\theta+2e\sin(\theta-\theta_p),
\end{equation}
where $\theta$ is measured from the equator, and $\theta_p$ is the phase angle defining the direction of perigee of the Keplerian ellipse on which the perturbations of $J_2$ are superimposed.\\\\
Using the usual unit vectors, $\vec{\hat{i}}$, $\vec{\hat{j}}$, $\vec{\hat{k}}$, with $\vec{\hat{i}}$ vertically upward, $\vec{\hat{j}}$ forward and $\vec{\hat{k}}$ perpendicular to the precessing plane, the actual position to the first order can be written as
\begin{equation}
\vec{r}=(\bar{r}+\delta r)\vec{\hat{i}}+\bar{r}\delta\theta\vec{\hat{j}}.
\end{equation}
And then the angular velocity of the unit vector frame is
\begin{equation}\label{3.81}
\vec{\omega_F}=\dot{\theta}\vec{\hat{k}}+\lambda_A\vec{\hat{N}},
\end{equation}
where $\vec{\hat{N}}=\vec{\hat{i}}\sin i \sin \theta+\vec{\hat{j}}\sin i\cos \theta+\vec{\hat{K}}\cos i$ is a unit vector directed northward along the earth's polar axis. As a result, the actual velocity relative to the Earth's centre is 
\begin{equation}
\vec{v}=\delta\dot{r}\vec{\hat{i}}+\bar{r}\delta\dot{\theta}\vec{\hat{j}}+\vec{\omega_F}\times\vec{r}.
\end{equation}
Using \eqref{3.78}, \eqref{3.79} and \eqref{3.81} in the above equation, we will get to the first order, the following expression
\begin{multline}\label{3.83}
\vec{v}=(\delta\dot{r}-\bar{r}\dot{\theta}\delta\theta)\vec{\hat{i}}+\left\{\dot{\theta}\left[1-\frac{3}{2}J_2\left(\frac{R_e}{\bar{r}}\right)^2\cos^2 i\right](\bar{r}+\delta r)+\bar{r}\delta\dot{\theta}\right\}\vec{\hat{j}}+\frac{3}{4}\left\{J_2\left(\frac{R_e}{\bar{r}}\right)^2\bar{r}\dot{\theta}\sin 2i\cos \theta\right\}\vec{\hat{k}}.
\end{multline}
The GR gravitational acceleration, $\vec{g}$, up to second order Legendre polynomial which can be found in literature by taking the multipole expansion of the potential and then taking the gradient is given by
\begin{equation}
\vec{g}=-GM\frac{\vec{r}}{r^3}+\vec{\nabla}\left\{\frac{GMJ_2R_e^2}{r^3}\left[\frac{1}{2}-\frac{3}{2}\frac{(\vec{r}\cdot\hat{N})^2}{r^2}\right]\right\},
\end{equation}
which then if expanded, gives to the first order the following result
\begin{equation}\label{3.85}
\begin{aligned}
\vec{g}=&-\frac{GM}{\bar{r}^2}\left\{1-2\frac{\delta r}{\bar{r}}+\frac{3}{2}J_2\left(\frac{R_e}{\bar{r}}\right)^2\left[1-\frac{3}{2}\sin^2 i(1-\cos 2\theta)\right]\right\}\vec{\hat{i}}\\&-\frac{GM}{\bar{r}^2}\left\{\delta\theta+\frac{3}{2}J_2\left(\frac{R_e}{\bar{r}}\right)^2 \sin^2 i\sin 2\theta\right\}\vec{\hat{j}}\\&-\frac{3}{2}\frac{GM}{\bar{r}^2}\left\{J_2\left(\frac{R_e}{\bar{r}}\right)^2\sin 2i \sin\theta\right\}\vec{\hat{k}}.
\end{aligned}
\end{equation}
Now, we require that equation \eqref{3.83} and \eqref{3.85} to be substituted in \eqref{3.75}, in order to get the corrected geodetic angular frequency for a oblate Earth.\\\\
To the first order, the result can be given by
\begin{equation}
\begin{split}
\vec{\Omega}_{geodetic}= & \left[\frac{3}{2c^2}-\frac{1}{6c^2}\exp(-\Upsilon r)\right]\frac{(GM)^{\frac{3}{2}}}{\bar{r}^{\frac{5}{2}}} \\
 &\left\{\vec{\hat{k}}\left[1+J_2\left(\frac{R_e}{\bar{r}^2}
\right)^2\left(\frac{9}{4}-\frac{27}{8}\sin^2i+\frac{9}{4}\sin^2i\cos 2\theta\right)+\frac{1}{\theta}\delta\dot{\theta}-\frac{1}{\bar{r}}\delta r\right] \right.\\
&\quad \left.{} -\frac{3}{4}J_2\left(\frac{R_e}{\bar{r}}\right)^2(2\vec{\hat{i}}\sin\theta+\vec{\hat{j}}\cos\theta)\sin2i\right\},
\end{split}
\end{equation}
Now $2\vec{\hat{i}}\sin\theta+\vec{\hat{j}}\cos\theta=\dfrac{3}{2}\vec{\hat{B}}-\dfrac{1}{2}\vec{\hat{B}}\cos2\theta+\dfrac{1}{2}\vec{\hat{A}}\sin2\theta$, where $\vec{\hat{A}}$ is the unit vector along the upward vertical at the ascending node and $\vec{\hat{B}}$ is the unit vector along a direction $(90)^0$ ahead of $\vec{\hat{A}}$ in the precessing plane. \\\\
The averages per orbit of $J_2\vec{\hat{B}}\cos2\theta$, $J_2\vec{\hat{A}}\sin2\theta$, $J_2\vec{\hat{k}}\cos2\theta$, $\vec{\hat{k}}\delta\dot{\theta}$ and $\vec{\hat{k}}\delta r$ are of second order. So, till the first order, our expression for the geodetic angular frequency becomes
\begin{multline}
\left\langle\vec{\Omega}\right\rangle_{geodetic}=\left[\frac{3}{2c^2}-\frac{1}{6c^2}\exp(-\Upsilon r)\right]\frac{(GM)^{\frac{3}{2}}}{\bar{r}^{\frac{5}{2}}}\\\left\{\left\langle\vec{\hat{k}}\right\rangle\left[1+J_2\left(\frac{R_e}{\bar{r}}\right)^2\left(\frac{9}{4}-\frac{27}{8}\sin^2i\right)\right]-\frac{9}{8}\left\langle\vec{\hat{B}}\right\rangle J_2\left(\frac{R_e}{\bar{r}}\right)^2\sin2i\right\},
\end{multline}
where the expressions $\left\langle\vec{\hat{k}}\right\rangle$ and $\left\langle\vec{\hat{B}}\right\rangle$ means that $\vec{\hat{k}}$ and $\vec{\hat{B}}$ each change by small amount because of nodal regression over the course of an orbit .\\\\
Since the satellite was inserted in the polar orbit, we consider only the polar orbit result for the geodetic angular frequency by substituting $i=(90)^0$ in the above equation. Hence, we get
\begin{equation}
\left\langle\vec{\Omega}\right\rangle_{geodetic}|_{polar}=\left[\frac{3}{2c^2}-\frac{1}{6c^2}\exp(-\Upsilon r)\right]\frac{(GM)^{\frac{3}{2}}}{\bar{r}^{\frac{5}{2}}}\left[1-\frac{9}{8}J_2\left(\frac{R_e}{\bar{r}}\right)^2\right]\vec{\hat{k}}.
\end{equation}
\subsection{\textbf{Evaluation of the terms of geodetic frequency}}\label{rudra}
In our calculation, we have the following quantities
\begin{equation}
\Phi=\frac{GM}{r},
\end{equation}
\begin{equation}
\vec{A}=\frac{G}{r^3}\left(\vec{r}\times\vec{J}\right).
\end{equation}
Let us first deal with the second term of equation \eqref{3.50} since it is trivial and we get
\begin{equation}\label{3.55}
\frac{3}{2}\vec{v}\times\vec{\nabla}\Phi=\frac{3}{2}\vec{v}\times\left(-\frac{GM}{r^2}\hat{r}\right)=\frac{3GM}{2r^3}\left(\vec{r}\times\vec{v}\right).
\end{equation}
Now, let's calculate the first term which is given as
\begin{equation}\label{3.56}
\vec{\nabla}\times\vec{A}=G\vec{\nabla}\times\left(\frac{\vec{r}}{r^3}\times\vec{J}\right),
\end{equation}
where
\begin{equation}
\vec{\nabla}\times\left(\frac{\vec{r}}{r^3}\times\vec{J}\right)=\frac{\vec{r}}{r^3}(\vec{\nabla}\cdot\vec{J})-\vec{J}\left(\vec{\nabla}\cdot\frac{\vec{r}}{r^3}\right)+(\vec{J}\cdot\vec{\nabla})\frac{\vec{r}}{r^3}-\left(\frac{\vec{r}}{r^3}\cdot\vec{\nabla}\right)\vec{J}.
\end{equation}
So, the equation \eqref{3.56} gives
\begin{equation}\label{3.65}
\vec{\nabla}\times\vec{A}=G\frac{\vec{J}}{r^3}-3G\frac{\vec{r}(\vec{r}\cdot\vec{J})}{r^5}.
\end{equation}
Now, let's calculate the last term in equation \eqref{3.52}
\begin{equation}
\vec{\nabla}(\Phi F)\times\vec{v}=\vec{\nabla}\left[\Phi\frac{\exp{(-\Upsilon r)}}{3}\right]\times\vec{v},
\end{equation}
where
\begin{equation}
\vec{\nabla}\left[\Phi\frac{\exp(-\Upsilon r)}{3}\right]=-GM\frac{\vec{r}}{r^3}\frac{\exp(-\Upsilon r)}{3}-\frac{GM\Upsilon}{3}\frac{\vec{r}}{r^2}\exp(-\Upsilon r).
\end{equation}
So, we have
\begin{equation}\label{3.68}
\vec{\nabla}(\Phi F)\times\vec{v}=\left[-\frac{GM}{r^2}\left(\frac{1}{r}\frac{\exp(-\Upsilon r)}{3}+\frac{\Upsilon}{3}\exp(-\Upsilon r)\right)\right]\left[\vec{r}\times\vec{v}\right].
\end{equation}
Hence, using \eqref{3.55},\eqref{3.65} and \eqref{3.68} in \eqref{3.52} gives
\begin{equation}
\vec{\Omega_{f(R)}}=3G\frac{\vec{r}(\vec{r}\cdot\vec{J})}{r^5}-G\frac{\vec{J}}{r^3}-\frac{3GM}{2r^3}(\vec{r}\times\vec{v})+\frac{1}{2}\left[\frac{GM}{r^2}\left(\frac{1}{r}\frac{\exp(-\Upsilon r)}{3}+\frac{\Upsilon}{3}\exp(-\Upsilon r)\right)\right]\left[\vec{r}\times\vec{v}\right].
\end{equation}


\subsection{\textbf{Consistency check of the Gyroscopic precession in metric $f(R)$ vs Brans--Dicke}}
\label{app:2}
Let us first see how metric $f(R)$ can be framed as a Brans--Dicke class theory.\\\\
The action of metric $f(R)$ theory is
\begin{equation}
\mathcal{S}=\frac{1}{2\kappa}\int d^4x\sqrt{-g}f(R)+S_M(g_{\mu\nu},\psi),
\end{equation}
where the symbols have their usual meanings and $\kappa=\dfrac{8\pi G}{c^4}$. One can introduce an auxiliary field $\chi$ and write a dynamically equivalent action \cite{teyssandier1983cauchy}:
\begin{equation}
\mathcal{S}=\frac{1}{2\kappa}\int d^4x \sqrt{-g}[f(\chi)+f'(\chi)(R-\chi)] +S_M(g_{\mu\nu},\psi).
\end{equation}
Variation with respect to $\chi$ results in
\begin{equation}
\chi=R,
\end{equation}
iff $f''(\chi)\neq 0$.\\\\
Redefining field $\chi$ by $\Phi=f'(\chi)$ and setting
\begin{equation}
V(\Phi)=\chi (\Phi)\Phi -f(\chi(\Phi)),
\end{equation}
the action becomes
\begin{equation}\label{2.147}
\mathcal{S}=\frac{1}{2\kappa}\int d^4x\sqrt{-g}(\Phi R-V(\Phi))+S_M(g_{\mu\nu},\psi).
\end{equation}
So, we immediately observe that metric $f(R)$ is the action of Brans--Dicke theory with $\omega=0$ or with a vanishing kinetic term, and hence we get from \eqref{3.110}
\begin{equation}\label{3.111}
\begin{aligned}
\vec{\Omega_{f(R)}}=&3\frac{\vec{r}(\vec{r}\cdot\vec{J})}{\phi_0 r^5}-\frac{\vec{J}}{\phi_0r^3}-\frac{3}{2}\frac{M}{\phi_0 r^3}(\vec{r}\times\vec{v})+ \left[\frac{M}{\phi_0 r^3}\frac{\exp(-m_s r)}{3}\vec{r}+\frac{M}{\phi_0 r^2}m_s\frac{\exp(-m_s r)}{3}\vec{r}\right](\vec{r}\times\vec{v})\\&+\frac{11V_0}{144\phi_0}(\vec{r}\times\vec{v}).
\end{aligned}
\end{equation}
This expression, in principle, can be used to find the gyroscopic precession frequency induced by any arbitrary function of $f(R)$ provided that we chose a metric variation. 

\subsection{\textbf{Gyroscopic precession in metric $f(R)$ ($f$ analytic)}}\label{laxmi}
First, let us evaluate $V_0$, using the assumption that $f(R)$ is an analytic function and can be expanded as 
\begin{equation}
f(R)=a_0+a_1 R+\frac{a_2}{2!}R^2...
\end{equation}
Since $\chi=R $ and $\Phi=f'(\chi)=f'(R)$, we have for the expression of potential
\begin{equation}\label{2.184}
\begin{split}
V(\phi) &=R(\phi)f'(R)-f(R)\\
&=R^{(1)}(1+a_2 R^{(1)})-(R^{(1)}+a_2 {R^{(1)}}^2)\\
&=\frac{1}{2}a_2 {R^{(1)}}^2(\phi),
\end{split}
\end{equation}
where we have introduced only the relevant terms of $f(R)$ \eqref{377}\eqref{378}. Now, since $R^{(1)}=\partial_{\mu}\partial_{\rho}h^{\rho\mu}-\Box h$ \eqref{1.112} is at least second order derivative of the metric perturbation, we can neglect it in the expression of precession frequency which is a linear one and hence end up with the following expression 
\begin{equation}\label{3.114}
\vec{\Omega_{f(R)}}=3\frac{\vec{r}(\vec{r}\cdot\vec{J})}{\phi_0 r^5}-\frac{\vec{J}}{\phi_0r^3}-\frac{3}{2}\frac{M}{\phi_0 r^3}(\vec{r}\times\vec{v})+ \left[\frac{M}{\phi_0 r^3}\frac{\exp(-m_s r)}{3}\vec{r}+\frac{M}{\phi_0 r^2}m_s\frac{\exp(-m_s r)}{3}\vec{r}\right](\vec{r}\times\vec{v}).
\end{equation}
Now let us compare it with the expression of precession frequency that we obtained from the linearised theory directly which is given by \eqref{3.69}
\begin{equation}
\vec{\Omega_{f(R)}}= 3G\frac{\vec{r}(\vec{r}\cdot\vec{J})}{r^5}-G\frac{\vec{J}}{r^3}-\frac{3GM}{2r^3}(\vec{r}\times\vec{v}) +\frac{1}{2}\left[\frac{GM}{r^2}\left(\frac{1}{r}\frac{\exp(-\Upsilon r)}{3}+\frac{\Upsilon}{3}\exp(-\Upsilon r)\right)\right]\left[\vec{r}\times\vec{v}\right].
\end{equation}
Immediately we observe that for the two approaches to be equivalent, we must require that $m_s=\Upsilon$ (Keep in mind that in the expression for gyroscope precession frequency for Brans--Dicke with a potential, the gravitational constant, $G$, is taken to be unity).
So, from \eqref{2.162}, we must have the following
\begin{equation}\label{2.187}
\frac{\phi_0}{2\omega+3}\frac{d^2V(\phi_0)}{d\phi^2}=m_s^2=\Upsilon^2.
\end{equation}
From \eqref{2.184} and using the quantities \eqref{377}\eqref{378}
\begin{align}\label{379}
f(R) &=R^{(1)}+\frac{a_2}{2!}{R^{(1)}}^2, & f'(R) &=1+a_2 R^{(1)},
\end{align}
where $R^{(1)}$ is the linearised Ricci Scalar, we can prove the expression \eqref{2.187}. Indeed, one can show that
\begin{equation}\label{2.194}
m_s^2=\frac{\phi_0}{3}\left(a_2\left(\frac{dR^{(1)}(\phi)}{d\phi}\right)^2+R^{(1)}(\phi)\frac{d^2R^{(1)}(\phi)}{d\phi^2}\right)\bigg\rvert_{\phi_0},
\end{equation}
and given that
\begin{align}
\phi =f'(\chi)=f'(R),
\end{align}
while from \eqref{378}, $f'(R)=1+a_2 {R^{(1)}}$, we have
\begin{equation}\label{2.196}
R^{(1)}=\frac{\phi-1}{a_2}.
\end{equation}
Making use of \eqref{2.196} in \eqref{2.194}, the second term in \eqref{2.194} goes to zero being the second derivative and we are left with
\begin{equation}\label{2.197}
m_s^2=\frac{\phi_0}{3a_2}.
\end{equation}
Now, for arguments of stability, $\phi_0$ has to be the minimum of the potential. To find the minimum of \eqref{2.184}, we have 
\begin{equation}
\frac{dV}{d\phi}=a_2 R^{(1)}\frac{dR^{(1)}}{d\phi}\bigg\rvert_{\phi_0}=0.
\end{equation}
Again, from \eqref{2.196}, the above expression can be written as 
\begin{equation}
a_2\left(\frac{\phi-1}{a_2}\right)\frac{1}{a_2}\bigg\rvert_{\phi_0}=0,
\end{equation}
which finally gives us the minimum value $\phi_0$, i.e., 
\begin{equation}
\phi_0=1.
\end{equation}
Hence, after substituting the found minimum value in \eqref{2.197}, the following result is derived
\begin{equation}\label{2.201}
m_s^2=\bigg\lvert\frac{1}{3 a_2}\bigg\rvert.
\end{equation}
We have argued that $a_2$ needs to be less than zero for $\Upsilon^2$ to be positive (since $\Upsilon^2=-\dfrac{1}{3a_2}$) which is required for physical solutions of the massive KG equation for the scalar mode. Also, in this theory, we remember that $m_s^2$ is positive as well (see \eqref{2.162}). Hence, we put a minus sign in \eqref{2.201} and get
\begin{equation}
m_s^2=-\frac{1}{3 a_2},
\end{equation}
so proving equation \eqref{2.187}. This demonstrates that the two approaches, one in which we considered $f(R)$ to be analytic \textit{a priori} and the second one, where we considered a generalised Brans--Dicke theory and searched for an expression of gyroscope precession frequency for metric $f(R)$ are equivalent. This is also a self--consistent non--trivial sanity check for both the approaches.\\\\
The equation \eqref{3.114} also reduces to that of GR, since $m_s=\Upsilon$ and $\phi_0=1$ as shown already and in the GR limit, $\Upsilon\rightarrow\infty$, i.e., $a_2\rightarrow 0$ and we just remain with the first three terms in \eqref{3.114} which is the GR result.

\bibliographystyle{spphys}       


\end{document}